\documentstyle[amssymb]{elsart}

\begin{document}

\input epsf

\begin{frontmatter}
\title{Possible Existence of an Extraordinary Phase in the Driven Lattice Gas}
\author{R. K. P. Zia, L. B. Shaw and B. Schmittmann} 
\address{Center for Stochastic Processes in Science and Engineering\\ 
Physics Department\\ 
Virginia Polytechnic Institute and State University \\ 
Blacksburg, VA, 24061-0435 USA} 
\begin{abstract}
We report recent simulation results which might indicate the existence of a 
new low-temperature ``phase" in an Ising lattice gas, driven into 
a non-equilibrium steady state by an external field. It appears that this 
``phase", characterized by multiple-strip configurations, is selected 
when square systems are used to approach the thermodynamic limit. We 
propose a quantitative criterion for the existence of such a ``phase".  
If confirmed, its observation may resolve a long-standing controversy 
over the critical properties of the driven Ising lattice gas.   
\end{abstract} 
\end{frontmatter}

\section{Introduction}

Nearly two decades ago, Katz, Lebowitz and Spohn introduced a seemingly
trivial generalization of the Ising model: a lattice gas driven far from
equilibrium by an external ``electric'' field, $E$ \cite{KLS}. The
motivations are twofold: the statistical mechanics of non-equilibrium
systems and the physics of fast ionic conductors \cite{FIC}. Since then,
this model has provided a rich variety of surprises, showing how our
intuition built on the foundations of equilibrium statistical mechanics
fails when applied to non-equilibrium stationary states. Though some of its
unusual properties are now understood, it continues to offer new surprises 
\cite{DL17}. In this paper, we report recent simulation results which may be
signals of a novel ``phase''. If this new state is established, it would
lead to a resolution of a long standing controversy concerning the critical
properties. In the next section, we briefly describe the model and the
disagreement between Marro, et. al. \cite{Marro} and the Leung-Wang team\cite
{KTLJSW}. Since the methods used in these two approaches are quite distinct,
we explore, in Section III, the possibility that the discrepancies may lie
in the existence of different types of ordering. In particular we introduce
a new parameter which distinguishes the two low temperature ``phases''. The
systems used in our simulations are simple generalizations of the standard
model -- to include anisotropic interparticle {\em interactions} -- so that
the presence of the novel phase is more pronounced even for small lattices.
The final section is devoted to some concluding remarks.

\section{The driven lattice gas and its critical properties}

The driven lattice gas is based on the Ising model \cite{Ising,YangLee} with
attractive nearest-neighbor interactions. In the spin language, it is a
ferromagnetic Ising model with {\em biased} spin-exchange \cite{Kawasaki}
dynamics. Since both the Onsager solution \cite{Onsager} and most
simulations concern a system in two dimensions ($d=2$), we give a brief
description of that model only. Each of the $L_{x}\times L_{y}$ sites of a
square lattice (on a torus, i.e., with periodic boundary conditions: PBC)
may be occupied by a particle or left vacant. A configuration of our system
is specified by the occupation numbers $\{n_{i}\}$, where $i$ is a site
label and $n$ is either 1 or 0. The interparticle attraction is modeled by
the usual Hamiltonian: ${\cal H}=-4J\sum_{<i,j>}n_{i}n_{j}$, where $<i,j>$
are nearest-neighbor sites and $J>0$. In thermal equilibrium, a half filled
system undergoes a second order phase transition, in the thermodynamic
limit, at the Onsager temperature $T_{O}=(2.2692..)J/k_{B}$. To perform
Monte Carlo simulations for this system, particles are allowed to hop to
vacant nearest neighbor sites with probability $\min [1,e^{-\Delta {\cal H}%
/k_{B}T}]$, where $\Delta {\cal H}$ is the change in ${\cal H}$ after the
particle-hole exchange. The deceptively simple modification introduced in 
\cite{KLS} is to bias the hops along, say, the $y$ axis, so that the new
rates are $\min [1,e^{-(\Delta {\cal H+}E\Delta y)/k_{B}T}]$. Locally, the
effect of $E$ is identical to that due to gravity. However, due to the PBC,
this modification cannot be accommodated by a (single-valued) Hamiltonian.
Instead, the system settles into a {\em non-equilibrium} steady state with a
non-vanishing global particle current. For temperatures above some finite
critical $T_{c}$, the particle density in this state is homogeneous, while
below $T_{c}$, the system displays phase segregation. Superficially, the
driven system behaves like the equilibrium Ising model. With deeper probing,
dramatic differences surface. For example, there is no question that its
critical properties fall {\em outside} the Ising universality class. For
recent reviews of this and other differences, we refer the interested reader
to \cite{DL17}.

Focusing on critical properties here, let us review a long standing
discrepancy between two sets of Monte Carlo results. While one set \cite
{KTLJSW} shows data entirely consistent with field theoretic renormalization
group predictions \cite{JSLC}, another group \cite{Marro} finds very
different results, for which no viable theoretical analysis exists \cite
{PsFT}. Technically, there are two serious differences between the
simulation methods. In this section, we conjecture that {\em different types
of ordering} may be responsible for the discrepancies. In view of the
presence of ``shape dependent'' thermodynamics in the {\em disordered} phase
of this system \cite{A+E}, we believe that the scenario pictured below is
quite plausible. Let us begin with a brief review of the controversy.

Consider the (untruncated!) two point correlation function (in spin
language): $G(x,y)\equiv \left\langle \left[ 2n(x,y)-1\right] \left[
2n(0,0)-1\right] \right\rangle $, and its Fourier transform: $S(k,p)\equiv
\sum_{x.y}G(x,y)e^{2\pi i\left( kx/L_x+py/L_y\right) }$, the structure
factor. Due to the conserved dynamics, $S(0,0)\equiv 0$, so that non-trivial
large distance properties are found in, e.g., $S(1,0)$ and $S(0,1)$. In
equilibrium systems with square geometries ($L_x=L_y=L$), isotropy dictates
that $S(1,0)=S(0,1)$, which approaches the static susceptibility in the
thermodynamic limit for $T>$ $T_O$. As $T\rightarrow $ $T_O,$ this quantity
diverges (as $L^{\gamma /\nu }$ for finite systems at $T_O$). Below $T_O$,
ordering is displayed as phase segregation, into a strip (of, say, a
particle rich region) aligned with either the $x$- (or $y$-) axis, so that $%
S(0,1)$ (or $S(1,0)$) $\rightarrow L^2$. For the {\em driven} system,
simulations \cite{KLS} first revealed a discontinuity singularity ($%
S(1,0)>S(0,1)$) in the disordered phase and showed further that, at low $T$,
phase segregation occurs {\em only} into ``vertical'' strips ($%
S(1,0)=O(L^2), $ $S(0,1)=O(1)$). Since these structure factors are related
to (the inverse of) diffusion co-efficients in a continuum theory, these
observations formed the basis for the study of a model where, first, the
diffusion co-efficients are anisotropic, and second, more importantly, {\em %
only one} of these co-efficients vanishes as $T\rightarrow $ $T_c$ \cite
{JSLC}. This approach is also consistent with subsequent simulations \cite
{Newton} showing that, while $S(1,0)$ diverges as $T\rightarrow $ $T_c$, $%
S(0,1)$ remains $O(1)$ throughout the transition. A serious consequence of
this ``strongly anisotropic'' behavior is that parallel and transverse
lengths and momenta scale with different powers ($p_y\sim p_x^{1+\Delta }$, $%
\Delta =1$ in mean field theory) in the critical region. Analogues of this
unusual property exist in equilibrium systems \cite{anisoEQ}. When
fluctuations are taken into account via renormalization group methods \cite
{JSLC}, the upper critical dimension ($d_c$) is found to be $5$, while
anomalous scaling modifies $\Delta $ to $2$ in $d=2.$ A more subtle result
is that one set of exponents retains the classical values.

Now, in order to /measure/ critical properties through simulations, with
some accuracy, it is crucial to exploit finite size scaling methods \cite
{FSS}. Systems with various $L$'s are used, and the quality of data collapse
serves as an indicator of the presence of scaling. Given the highly
anisotropic scaling of lengths, the analysis would be much simplified if
samples obeying $L_{y}\sim L_{x}^{3}$ are used. This route is pioneered and
followed in one of the studies \cite{KTLJSW}. More importantly, the order
parameter used is closely related to $S(1,0)$, so that it is sensitive to
the presence of a {\em single} domain (i.e., one vertical strip, regardless
of the system size). Also measured are cumulant ratios associated with $%
S(1,0)$. {\em All} data thus obtained display good collapse, using a set of
exponents that is entirely consistent with those predicted by the
renormalization group \cite{JSLC}. In another set of studies \cite{Marro},
only square samples are used. Furthermore, the order parameter used is

\begin{equation}
m\equiv \sqrt{\left\langle M_{y}^{2}\right\rangle -\left\langle
M_{x}^{2}\right\rangle }  \label{m}
\end{equation}
where $M_{x(y)}^{2}\equiv L^{-2}\sum_{y(x)}\left[ \sum_{x(y)}\left\{
2n_{i}-1\right\} \right] ^{2}$. Note that, {\em if} the ordered phase
consists of a single vertical strip, then $m$ provides information about its
average density. However, configurations with vertically aligned {\em %
multiple strips }also contribute to $m.$ The conclusion is that, e.g., $%
\beta \simeq \frac{1}{4}$, contradicting the field theory results \cite
{Marro}.

\section{A possible novel phase}

It should be quite alarming that two sets of simulations lead to drastically
different conclusions, unless we allow the possibility that the ``ordered''
states in the two studies are {\em not} the same. Considering $m$ is {\em not%
} sensitive to the presence of multiple strips, we conjecture the following.

\begin{itemize}
\item  In the thermodynamic limit, the low temperature state in systems with
square geometry ($L_x=L_y$) is characterized by multiple strips, with a
nontrivial distributions of widths.
\end{itemize}

In other words, we believe that there is no long range order
(which would lead to a single-strip state) while strip-strip correlations
are controlled by a finite correlation length. For short, we will use a more
picturesque term: ``stringy phase.''

%%%%%%%%%%%%%%%   Fig 1  %%%%%%%%%%%%
\begin{figure}[tbph]
\hspace*{2.5cm} 
\epsfxsize=5cm \epsfbox{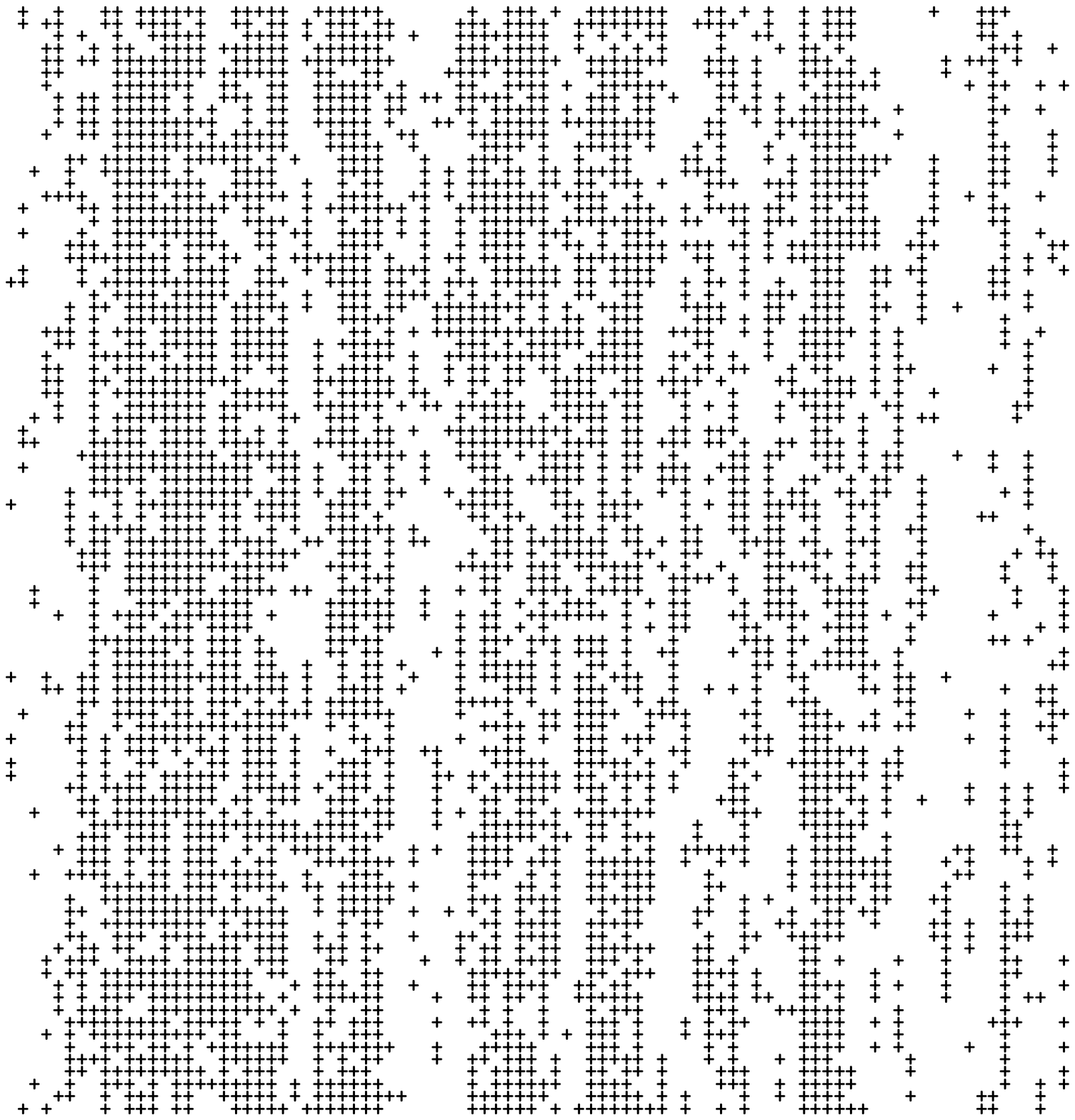}
\vspace*{-6.68cm}
\end{figure}

\begin{figure}[tbph]
\hspace*{8cm}
\epsfxsize=5cm \epsfbox{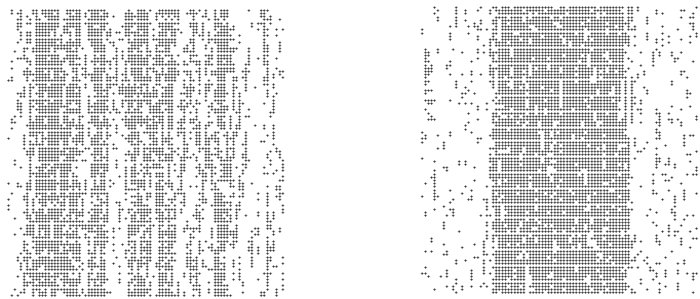}
\vspace*{-1.8cm}
\caption{A typical ``stringy'' configuration vs. a typical single strip one. 
(See text for system parameters.)}
\end{figure}
%%%%%%%%%%%%%%%%%%%%%%%%%%%%%%%%%%%%%%%

Significantly, the existence and
behavior of such configurations were first noted in \cite{Marro}. Indeed,
these authors wrote: ``The system was never seen to escape from (these
multiple-strip) states... and we had to manipulate some system
configurations for $L=50$ to create artificial one-strip states...''. For $%
L\geq 100$ systems, they were ``unable to observe the decay (of these
multiple-strip states) toward one-strip states...'' The conclusion was that
multiple states are ``metastable'' with exceedingly long life times.

Another source for our conjecture comes from simulation studies of a driven
lattice gas with anisotropic interparticle {\em interactions} ($J_x\neq J_y$)
\cite{LBS}. Though typical configurations are disordered/ordered(single strip) 
for high/low temperatures, there is a significant range of $T$ where the 
system appears to be ``stringy.'' In Fig.1, we show such a configuration at 
$T=0.80$, found in a $90\times 90$ lattice with $3J_x=J_y/3=J$, driven
with 
saturation $E$. (Here, as in the following, $T$ is given in units of $T_O$.) 
In contrast, the single strip configuration was obtained with $T=0.70$. 
To be more precise, 
the ``stringy'' states may be characterized by two properties: (a) the
densities in each column (i.e., along the drive) are bi-modally distributed,
yet (b) $S(1,0)$ remains small \cite{LBSMS}. An example is provided in Fig.2. 
While the column densities begin to develop a
double peak around $T\sim 0.90$, the structure factor displays an
inflection
point at $T\sim 0.70$ (which is also where the {\em fluctuations} in
$S(1,0)$
reach a maximum). Note that bi-modal column densities correspond to $m>0$.
Yet, in this range ($0.75\lesssim T\lesssim 0.95$), the typical configurations 
are far from being single-stripped. To show that multiple-strip states are
not simply metastable, we ran simulations starting from a completely ordered
(single-strip) state. Unlike in the previous study \cite{Marro}, we {\em do }%
observe the split-up into multiple strip 
states\footnote{Figures in the Marro studies 
(e.g., Fig 2.7 in \cite{MDbook}) show typical
configurations of $T$ well below the transition value. It is possible that
the failure to observe break-up/re-merging is due to $T$ being too low (for
the sizes studied).}. 
In Fig.3, we show a time trace of 3500 measurements
(700K MCS) of $S(k,0)$, $k=1,2,3$, which are sensitive to the presence of 
$k$-strip domains. Note that $S(1,0)$ is often smaller than $S(2,0)$ and
sometimes even smaller than $S(3,0)$, as shown in the inset. As a check, we
made three separate runs of this length and observed the same type of
behavior. Our conclusion is that, despite being started in a completely
ordered state, the system evolves towards a ``stringy'' state. In other
words, the probability to find the system in multiple strip configurations
can be considerable.

%%%%%%%%%%%%%%%   Fig 2 and 3   %%%%%%%%%%%%
\begin{figure}[tbph]
\hspace*{-1.2cm} \epsfxsize=17cm \epsfbox{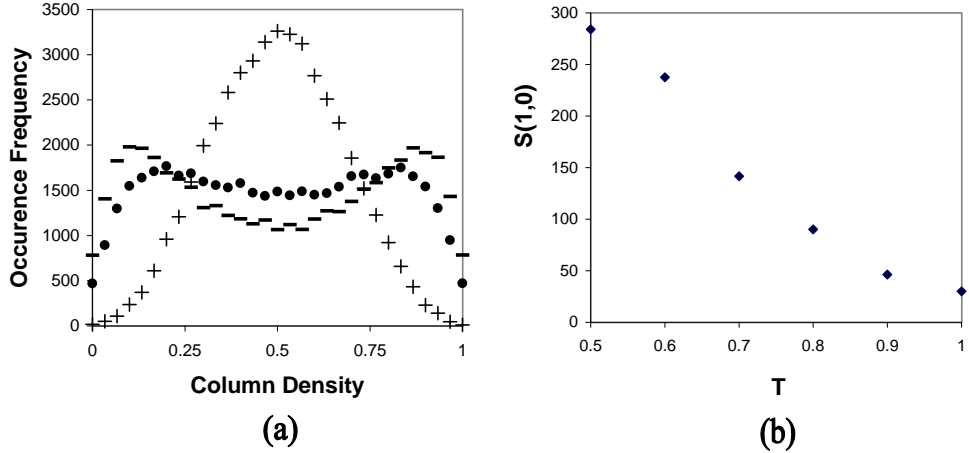} \vspace*{-6.5cm}
\caption{(a) Histogram for column densities at $T=0.80$[$-$],
$0.90$[$\bullet $], 
$2.0$[$+$], and (b) $S(1,0)$ vs. $T$ in a $30\times 30$ lattice with 
$3J_x=J_y/3=J$, driven with saturation $E$.}
\end{figure}

\begin{figure}[tbph]

\hspace*{-1.2cm} %\vspace*{3cm}
\epsfxsize=17cm \epsfbox{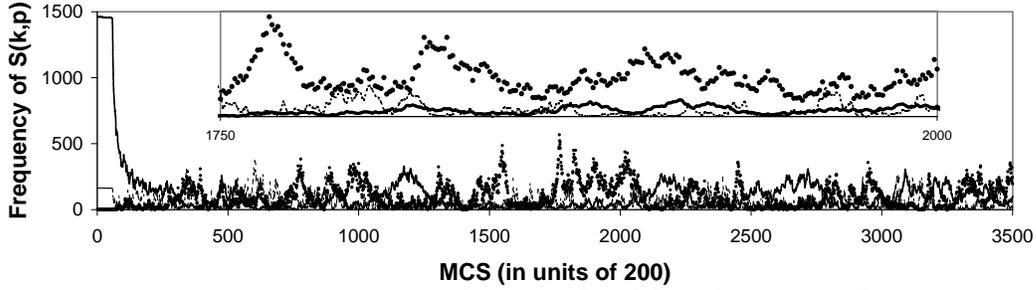} \vspace*{-8cm}
\caption{Time trace of $S(k,0)$, $k=1$ [solid line], $2$ [$\bullet $], $3$,
[dashed line], for a $60\times 60$ lattice at $T=0.85$, driven with
saturation $E$. Inset: magnified view of a portion.}
\end{figure}
%%%%%%%%%%%%%%%%%%%%%%%%%%%%%%%%%%%%%%%%%%

In order to make quantitative statements about the ``stringy state'', we
introduce the ratio:

\begin{equation}
{\cal R}\equiv L^{d}\frac{G(0,L/2)}{S(1,0)}  \label{R}
\end{equation}
for a system in $d$ dimensions. The numerator is a measure of ``long range
order'' along the $y$-axis, while the denominator detects the presence of a
single strip. In particular, we are interested in the behavior of ${\cal R}$%
, as $L\rightarrow \infty $, when the system is in various states.

Far in the disordered phase, the two-point correlation decays as $r^{-d}$ 
\cite{LRC}. Thus, $G(0,L/2)\rightarrow O(L^{-d})$, while $S(1,0)$ remains $%
O(1)$. On the other hand, deep in the ordered phase $G\rightarrow O(1)$
while $S\rightarrow O(L^d)$. As a result, we expect

\begin{equation}
{\cal R}\rightarrow O(1)  \label{far-from Tc}
\end{equation}
for both high and low temperatures. In between, a finite system makes a
transition, either through a ``stringy phase'' or directly into the ordinary
ordered state (as in the equilibrium case). In the latter scenario, the
results of the renormalization group analysis \cite{JSLC} should hold: $%
G(0,r)\rightarrow r^{-[d-2+\Delta ]/(1+\Delta )}$ and $S(k,0)\rightarrow
k^{-2}$. Thus, for simulations in $d=2$, we conclude that ${\cal R}$ {\em %
decreases} with $L:$

\begin{equation}
{\cal R}\rightarrow L^{-2/3}.  \label{DDS-Tc}
\end{equation}
In sharp contrast, for a{\em \ }stringy state, ordering has set in along $y$
(so that $G(0,L/2)\rightarrow O(1)$) but complete phase segregation in $x$
is yet to take hold (so that $S(1,0)\rightarrow O(1)$). The consequence is
an {\em increasing }${\cal R}:$

\begin{equation}
{\cal R}\rightarrow L^{d}.  \label{stringy}
\end{equation}
which is a function of $L.$ The drastic differences between these asymptotic
properties (\ref{far-from Tc}, and \ref{DDS-Tc} vs. \ref{stringy}) should be
detectable through simulations. As a final remark, we note that, for the 
{\em equilibrium} Ising model, ${\cal R}$ decreases exponentially as $%
L^{2}e^{-L/\xi }$ for $T>T_{c}$, while for $T\lesssim T_{c}$, ${\cal R}$ is $%
O(1)$.

Turning to Monte Carlo data, we find that the evidence in favor of stringy
states is quite strong, at least for the $L$'s we can access. In particular,
using models with $J_{x}=J/3$ and $J_{y}=3J$ under saturation drive, we
collected data from a number of systems with $L\in \left[ 10,90\right] $.
The runs are up to 800K MCS long, with both random and ordered initial
conditions. In Fig.4, we show a plot of ${\cal R}$ vs. $L^{2}$, for various
temperatures. For $T=0.85$ and $0.90$, this ratio appears to be
proportional
to $L^{2}$. Meanwhile, its behavior is consistent with ${\cal R}\rightarrow
O(1)$ at both higher and lower $T$'s. We believe that these results support
the presence of a ``stringy phase''.

%%%%%%%%%%%%%%%  Fig 4  %%%%%%%%%%%%%%%%%%%%

\begin{figure}[htbp]
\hspace*{-1.5cm} \epsfxsize=17cm \epsfbox{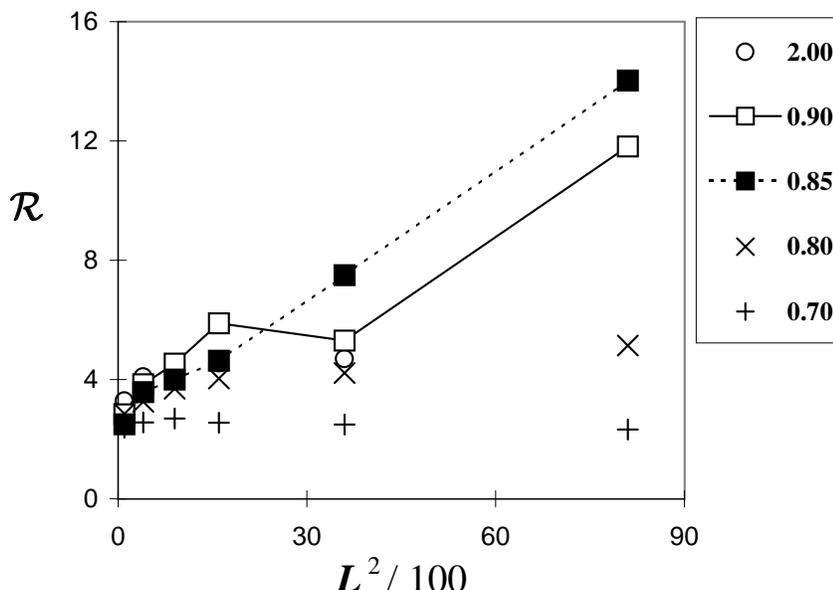} 
\vspace*{-14cm}
\caption{Correlation-Structure Factor ratio ${\cal R}$ for systems with $%
\alpha =3$ and saturation drive, plotted against $L^2.$ The legend refers to
the temperatures of the various runs.}
\end{figure}

%%%%%%%%%%%%%%%%%%%%%%%%%%%%%%%%%%%%%%%%%%%%

Of course, this ``phase'' may be just a mirage in the landscape of a giant
crossover. In the absence of analytic results, we can never be certain of
its existence, since all simulations are based on finite $L$. Nevertheless,
if future simulations continue to support ${\cal R}\propto O(L^2)$, then it
behooves us find a sound analytic basis for such a state. To conclude this
section, we conjecture further. As $L$ increases further, the stringy phase
will persist at lower and lower temperatures, so that it is the {\em only}
low temperature state, for {\em square} systems, in the limit $L\rightarrow
\infty $. However, if the thermodynamic limit is approached through systems
obeying $L_y\propto L_x^3$, the low temperature phase will consist of the
ordinary, {\em single-domain} state. If this conjecture turns out to be
true, then an avenue is opened for the resolution of the controversy over
critical properties.

\section{Concluding remarks}

In this short paper, we explored the possibility that the low temperature
phase of the driven lattice gas depends on the geometry of the system. For 
{\em square} systems, we conjecture that a new, ``stringy'' phase is the
only ordered state in the limit $L\rightarrow \infty $. On the other hand,
if an Ising-like, ordered state with a single domain is desired, then the
``unusual'' thermodynamic limit, with {\em rectangular }systems obeying $%
L_{\Vert }\propto L_{\bot }^3$, must be used. For finite square geometries,
we have some Monte Carlo evidence for the presence of a ``stringy'' state:
multiple strips of high/low density regions. The system appears to be
``ordered'' in the direction parallel to the drive, but the strips are of
varying widths and multiplicities. Specifically, as $T$ is lowered, the
disordered, homogeneous phase gives way to this multi-strip state for a
sizable range of $T$, before settling into the fully phase segregated,
single domain state at lower temperatures.

Many open questions, besides those mentioned above, naturally arise. More
quantitative analyses can be carried out, in order to better characterize
this ``stringy'' state. For example, we could measure the distribution of
strip-widths, or compile histograms for both the structure factors and the
correlations. Once a clearer picture emerges, assuming such a state still
exists, we should attempt to develop a reliable field theory. Certainly, the
order parameter of such a theory, $m$ in Eqn. (\ref{m}), is necessarily
quite different from the one for an equilibrium Ising lattice gas.
Hopefully, renormalization group methods can be applied, a nontrivial fixed
point can be obtained, and its associated critical properties can be
computed. Until then, to compare field-theoretic predictions with simulation
data from the Marro studies makes as much sense as ``comparing apples with
oranges.''

\section{Acknowledgments}

It is a pleasure to dedicate this article to Joel Lebowitz, a co-author of
this model. Over the years, we have benefitted from numerous conversations
with him, not to mention all the encouragements when we run into apparent
dead-ends. Also, we are grateful to J. Hager for help with simulations,
as well as G.L. Eyink and U.C. T\"{a}uber for illuminating
discussions. This research was supported in part by a grant from US National
Science Foundation through the Division of Materials Research and a
supplement from the Research Experience for Undergraduates program.

\end{document}